\documentclass{emulateapj}

\newcommand{\kms}{\mathrm{\;km\;s^{-1}}}

\shorttitle{A Dual AGN at $z=0.7$}
\shortauthors{Brian F. Gerke et al.}

\begin{document}

\title{The DEEP2 Galaxy Redshift Survey: AEGIS observations
  of a Dual AGN at $z=0.7$}

\author{Brian F. Gerke\altaffilmark{1},
Jeffrey A. Newman\altaffilmark{2,3}, 
Jennifer Lotz\altaffilmark{4},
Renbin Yan\altaffilmark{5}, 
P. Barmby\altaffilmark{6},
Alison L. Coil\altaffilmark{2,7},
Christopher J. Conselice\altaffilmark{8},
R.~J. Ivison\altaffilmark{9,10},
Lihwai Lin\altaffilmark{11},
David C. Koo\altaffilmark{12},
Kirpal Nandra\altaffilmark{13},
Samir Salim\altaffilmark{14},
Todd Small\altaffilmark{15},
Benjamin J. Weiner\altaffilmark{16}, 
Michael C. Cooper\altaffilmark{5},
Marc Davis\altaffilmark{1,5},
S. M. Faber\altaffilmark{12},
Puragra Guhathakurta\altaffilmark{12}}

\altaffiltext{1}{Dept. of Physics, U.C. Berkeley,
Berkeley, CA 94720}
\altaffiltext{2}{Hubble Fellow} 
\altaffiltext{3}{Lawrence Berkeley National Laboratory, 
Berkeley, CA 94720}
 \altaffiltext{4}{Leo Goldberg Fellow, NOAO, Tucson, AZ 85719}
\altaffiltext{5}{Dept. of Astronomy, U.C. Berkeley,
Berkeley, CA 94720}
\altaffiltext{6}{Harvard-Smithsonian Center for Astrophysics,
  Cambridge, MA 02138}
\altaffiltext{7}{Steward Observatory, U. of Arizona, Tucson,
  AZ 85721}
\altaffiltext{8}{School of Physics and Astronomy, U.  of
  Nottingham, University Park NG9\,2RD UK}
\altaffiltext{9}{UK Astronomy Technology Centre, Royal Observatory,
                 Blackford Hill, Edinburgh EH9\,3HJ, UK}
\altaffiltext{10}{Institute for Astronomy, U. of Edinburgh,
                 Blackford Hill, Edinburgh EH9\,3HJ, UK}
\altaffiltext{11}{Dept. of Physics, National Taiwan University, Taiwan}
\altaffiltext{12}{UCO/Lick Observatory, UCSC, Santa Cruz, CA 95064}
\altaffiltext{13}{Astrophysics Group, Imperial College, London SW7\,2BZ
  UK}
\altaffiltext{14}{Dept. of Physics and Astronomy, UCLA, Los Angeles, CA 90095} 
\altaffiltext{15}{Space Astrophysics, 405-47, Caltech, Pasadena, CA 91125} 
\altaffiltext{16}{Dept. of Astronomy, U. of Maryland, College Park,
  MD 20742}
\email{bgerke@astro.berkeley.edu}

\begin{abstract}
We present evidence for a dual Active Galactic Nucleus (AGN) within an
early-type galaxy at $z=0.709$ in the Extended Groth Strip. 
The galaxy lies on the red sequence, with absolute magnitude
$M_B=-21.0$ (AB, with $h=0.7$) and rest-frame color $U-B=1.38$.  Its
optical spectrum shows 
strong, double-peaked [\ion{O}{3}]  emission lines and weak H$\beta$
emission, with Seyfert-like line ratios.  The two
narrow peaks are separated by $630 \kms$ in 
velocity and arise from two distinct regions, spatially resolved in
the DEIMOS spectrum, with a projected physical separation of
$1.2$ kpc.  \emph{HST}/ACS imaging shows an 
early-type (E/S0) galaxy with hints of disturbed structure,
consistent with the remnant of a dissipationless merger.    
Multiwavelength photometric information from the AEGIS consortium
confirms the  
identification of a dust-obscured AGN in an early-type galaxy, with
detections in X-ray, optical, infrared and radio wavebands.
These data are most readily explained as a single galaxy harboring two
AGN---the first such system to be observed in an otherwise typical early-type  
galaxy. 

\end{abstract}



\section{Introduction}
\label{sec:introduction}

The standard picture of AGN emission---which
identifies the engines as  accreting supermassive
black holes (SMBHs)---is now well
established.  Also well established is the 
hierarchical paradigm for structure formation,
in which massive galaxies grow via a series of mergers of smaller
objects.  An obvious consequence of this picture is that some galaxies
should contain two SMBHs in their centers.  Such nuclear BH pairs will
initially be widely separated ($\ga 1$ kpc) ``dual'' SMBHs.  After
$\sim 100$ Myr they will
become true \emph{binary} SMBHs, gravitationally bound to one another
and with parsec-scale
separations; they may finally coalesce into a single central
SMBH on much longer timescales~\citep{BBR80}.
The behavior of SMBH binaries has 
significant implications for the formation of galactic cores (e.g.,
\citealt{MMRv02}), and their coalescence may be a
significant source of gravitational radiation in the universe; this
subject has received much theoretical study (for a review
see~\citealt{MM05}).    

Observational study of these objects has been far more limited,
however: only recently has direct detection of a true binary SMBH
been reported, in the LINER-like elliptical radio galaxy
0402+379 \citep{Maness04, Rodriguez06}.
In addition, there is only one unambiguous detection of a dual AGN in a single galaxy, 
from \emph{Chandra} studies of the merger remnant NGC 6240 \citep{Komossa03}, although
there are several examples of merging pairs of AGN hosts
(\emph{e.g.}, \citealt{Ballo04, Guainazzi05, Hudson06}), and there is a population of
close quasar pairs that are very unlikely to be lensed objects (\emph{e.g.} \citealt{Kochanek99, 
Junkkarinen01}).  Most of the merger examples were discovered in studies of merging 
late-type galaxies with significantly disturbed morphologies.
Recent work on galaxy evolution 
(e.g., \citealt{Faber05}) implies that significant stellar mass in
elliptical galaxies is built up in dissipationless 
mergers of early-type progenitors.
The remnants of such mergers exhibit only subtle visual
signatures \citep{vanD05}, suggesting 
that dual-SMBH systems should also occur in apparently normal 
early-types. 

The Extended Groth Strip is a subregion of 
the DEEP2 Galaxy redshift survey that is also the site of
intense, multiwavelength observational efforts by the All-wavelength
Extended Groth strip International 
Survey (AEGIS) team.  Full details of the dataset can be found in
\citet{AEGIS}.  In this Letter, we present evidence from AEGIS 
for a dual AGN at $z=0.709$ in the $R_{AB} = 22.6$ early-type galaxy
EGSD2 J142033.66+525917.5
The object was identified serendipitously in the
process of visually inspecting DEEP2 spectra to confirm galaxy
redshifts.  
Throughout this work we adopt a flat,
$\Lambda$CDM cosmology with $\Omega_M=0.3$ and $h=0.7$; all distances
are quoted in physical (not comoving) units.

\section{Observations}
\label{sec:results}

\begin{figure}
\centering
\includegraphics[width=0.7\linewidth, angle=90]{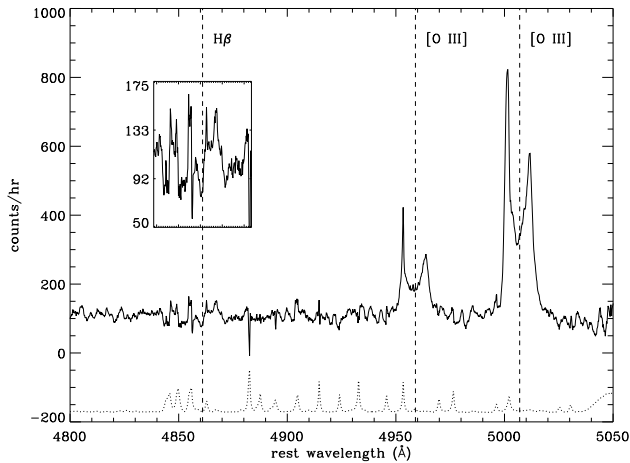}
\caption{A portion of the one-dimensional DEIMOS spectrum of the
  dual-AGN host galaxy (solid line). 
  Night-sky emission has been subtracted, and the spectrum has been
  weighted by its inverse variance and smoothed 
  with a ten-pixel-wide top-hat kernel.  The dotted line shows the
  Poisson uncertainty in the smoothed spectrum, scaled up by a
  factor of three and offset for clarity.
  The wavelength scale has been shifted to the host galaxy's rest frame
  (see text);
  dashed vertical lines show the expected wavelengths of [\ion{O}{3}]
  and 
  H$\beta$ emission. Each [\ion{O}{3}] line has two clear
  emission components 
  separated by   $630\kms$ in velocity.
  For clarity, the H$\beta$ emission is shown on a magnified vertical
  scale in the 
  inset; it is affected by noise from a night sky line.}
\label{fig:1dspec}
\end{figure}

\begin{figure}
\plotone{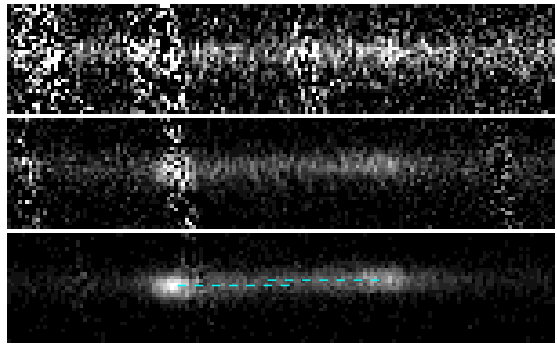}
\caption{Two-dimensional spectral emission on the DEIMOS detector in
  the vicinity of the emission lines seen in Fig.~\ref{fig:1dspec},
  after sky subtraction.
  In each panel, wavelength increases to the right, and spatial
  position along the slit (oriented at position angle $41^\circ$) is
  represented in the vertical direction.  From top to  
  bottom, the lines shown are H$\beta$ (partially obscured by
  night-sky noise), [\ion{O}{3}]$\lambda$4959, and 
  [\ion{O}{3}]$\lambda$5007.  Horizontal dashed lines in the bottom panel
  show the spatial centroids of the two emission line components,
  which are offset from each other by 1.5 DEIMOS pixels, or
  $0^{\prime\prime}.17$.  This corresponds to a projected
  separation of 
  1.2 kpc.} 
\label{fig:2dspec}
\end{figure}

Fig.~\ref{fig:1dspec} shows part of the one-dimensional DEEP2 
spectrum of this object, obtained using the DEIMOS spectrograph on the 
Keck II telescope.  
The most notable features of the spectrum are a pair of
strong, double-peaked emission lines, readily identifiable by their
wavelength ratios as [\ion{O}{3}] $\lambda$4959 and $\lambda$5007.
The two peaks of the $\lambda$5007 line  are separated by $17.8$\AA\ in 
wavelength.  The spectrum also has  H$\beta$ emission, which may be
double-peaked but is 
partially obscured by residual night-sky Poisson noise;  there are no
other emission lines, but the spectrum 
displays a numerous stellar continuum absorption features over the 
entire observed wavelength range (6750--8890\AA), most of which is
not shown here.  Fig.~\ref{fig:2dspec} shows the 
sky-subtracted two-dimensional DEIMOS spectrum of this object in the
regions around each of the emission lines seen in
Fig.~\ref{fig:1dspec}.   The spectrum exhibits two \emph{spatially offset} 
emission components: their (one-dimensional) spatial centroids (which
can be measured independently to an accuracy of $\sim 0.1$ pixels) 
are separated by 1.5 DEIMOS pixels, or $0^{\prime\prime}.17$ on the sky.  
The two components' spatial extents are consistent with the
$\sim 0^{\prime\prime}.6$   
seeing on the night of observation; they are not extended sources.

Because the emission lines in this spectrum have
significant velocity structure, we determined the systemic redshift of
the galaxy by fitting synthetic spectral templates
to the galaxy's stellar continuum emission, whose
absorption features do not exhibit such structure.  Following
\citet{Yan06}, we use the stellar population synthesis code of
\citet{BC03} to create
two templates, representing an old (7 Gyr) population and a young 
(0.3 Gyr) population, both with solar metallicity.  We then find the
redshift and the linear combination of these templates that best fit
the observed spectrum, excluding wide windows around the H$\beta$  
and [\ion{O}{3}] lines.  Only the old-population template proved
necessary for a good fit. 
This procedure yields a redshift of $z=0.709$ for the
host galaxy.  The K-correction algorithm of \citet{Willmer05} then
gives an absolute magnitude $M_B=-21.0$ (AB) and a rest-frame color
$U-B=1.38$---among the reddest in DEEP2, though typical of AGN
\citep{Nandra06}. 
We have shifted the wavelength scale of Fig.~\ref{fig:1dspec}
to the rest frame of the host; the expected wavelengths of the
emission lines are shown by vertical dashed lines.  The two 
emission peaks are located almost symmetrically with
respect to the host galaxy in velocity space.
In the host's rest frame, their velocity separation is $630 \kms$
and their projected physical separation is $1.2$ kpc.  

Having fit the stellar continuum, we may subtract
off this emission and consider the properties of the emission
lines alone.  First, the two
emission-line components have strongly asymmetric profiles.  The
emission lines seen in Fig.~\ref{fig:1dspec} are not well described
by the sum of two Gaussian curves;
there is significant emission bridging the two peaks.  
We therefore fit the spectrum with \emph{three} Gaussians by
iteratively 
fitting and subtracting first the blueward peak, then the redward, 
then
the bridging emission. This yields velocity dispersion measurements
of $54 \pm 2 \kms$ and $95 \pm 4 \kms$ for the blueward
and redward emission components, respectively.  The bridging emission
is too uniformly distributed in wavelength to be particularly well
described a Gaussian, but its nominal dispersion is $150\pm 8 \kms$,
and its peak 
is consistent with zero velocity offset.  The ratio of the
[\ion{O}{3}] $\lambda 5007$ and H$\beta$ line fluxes is commonly used to
distinguish Seyfert galaxies from other emission-line galaxies, with
Seyferts having $\lambda 5007$/H$\beta > 3$ in the local universe
(e.g., \citealt{Kauffmann03}); this criterion remains
valid for red galaxies at $z\sim 1$ (R.~Yan et al. in prep.). We have
measured the 
ratios for the two emission components in this red galaxy by summing
the 1-d spectrum over the measured $1\sigma$ velocity width of each
peak.  
This procedure gives ratios of $11.9\pm 0.4$ and $15.8 \pm 0.4$ for
the blueward and redward component, respectively,
placing both firmly in the Seyfert category.

\begin{figure}
\plotone{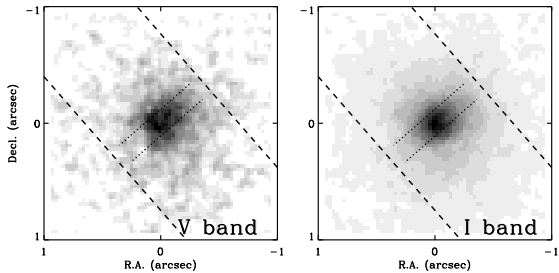}
\caption{
\emph{HST}/ACS images of the host galaxy in the F606W and F814W ($V$
and $I$) bands.  Both images are shown with a logarithmic intensity
scale and have 
been smoothed by a Gaussian kernel with a 1.5-pixel FWHM.  The
galaxy appears to be spheroidal, with a roughly circular profile, but
both images show subtle asymmetries, possibly indicating disturbed
structure.  Also shown 
are the edges of the DEEP2 DEIMOS
slit (dashed lines; approximately $50$\% of the full slit length is
shown) and the positions along the slit of the two 
emission-line components in Fig.~\ref{fig:2dspec} (dotted
lines).    
}
\label{fig:hst}
\end{figure}

We now turn to multiwavelength imaging from the AEGIS 
collaboration.
Of the available imaging data, only the
\emph{HST}/ACS images have the potential to resolve
the two emission regions;
we show ACS imaging in the F606W and F814W
(hereafter $V$ and $I$) bands in Fig.~\ref{fig:hst}.
Also shown are the
positions of the DEIMOS slit edges and the 
positions along the slit of the emission components seen in
Fig.~\ref{fig:2dspec}.  The $I$ band image shows an
early-type, spheroidal galaxy with a smooth profile, but the image
also exhibits 
subtle asymmetry, with a nucleus that is slightly off-center compared
to the  outer envelope; this may indicate recent merger activity.  The
$V$ band  image is less smooth and more obviously asymmetric.  It also 
shows an interesting horseshoe-shaped structure near its center on a
similar angular scale as the galaxy's two emission-line components;
this structure might be related to the emission regions.

We have run the GALFIT galaxy image modeling code \citep{Galfit}  on
the ACS images, and we find that both
radial profiles are described well by a \citet{Sersic68} profile, with
index n=2.14 for the $I$ band and n=1.7 for the $V$ band. 
The images' central concentrations are thus near the low end of the
elliptical range.  Also, the morphological parameters computed by
\citet{Lotz06} for the ACS imaging are consistent with 
expectations for bulge-dominated galaxies, but with a significantly
lower concentration and  
higher degree of substructure than are typical for such objects.
These results support the picture of an early-type galaxy with a
double nucleus.

\begin{figure}
\centering
\includegraphics[width=0.7\linewidth, angle=90]{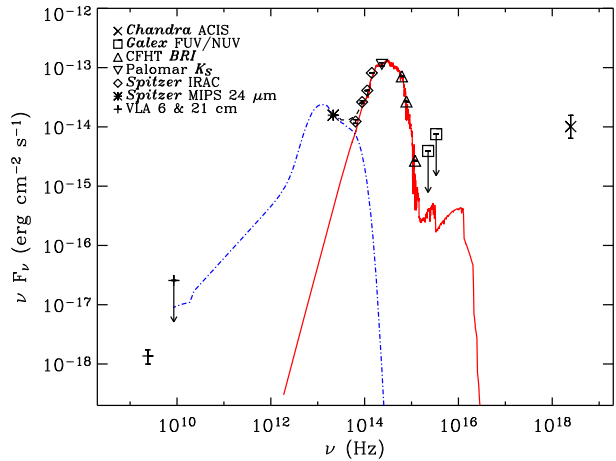}
\caption{Spectral energy distribution of the dual AGN and host galaxy,
from radio to X-ray.  Data points show fluxes (or upper limits)
measured by various
instruments, as noted in the legend.  All frequencies have been shifted
to the host galaxy's rest frame.  Also shown are a template spectrum
for an early-type galaxy (see text) 
normalized to the K-band flux (solid
line) and a simple model for infrared emission from a dust-obscured
AGN 
normalized to the MIPS 24 $\mu$m flux (dot-dashed line).  The dashed line
(visible at center) shows the sum of the dust and stellar models.} 
\label{fig:sed}
\end{figure}

In Fig.~\ref{fig:sed}, we compile a rest-frame SED 
from AEGIS observations. 
The object is detected at $>3\sigma$ significance in
\emph{Chandra}/ACIS imaging,
ground-based $BRIK$ photometry from CFHT and Palomar, all four
\emph{Spitzer} IRAC bands, the \emph{Spitzer} MIPS 24 $\mu$m band, and
at 21cm with the VLA.  We also show $3\sigma$ upper limits on the flux
in the \emph{GALEX} bands
and at 6cm.  

The measured SED is
well described by the simple model of an early-type galaxy with a
central, dust-obscured AGN. 
Plotted in Fig.~\ref{fig:sed} is a
synthetic SED, produced using the code of \citet{BC03}, representing a
galaxy with an old (7 Gyr), dust-free stellar population that formed
in a single 500 Myr burst; it has been normalized to the
measured K band flux of the host galaxy, and no further fitting has
been done in the figure.   To obtain more precision, we have applied
the SED template-fitting procedure of \citet{Salim05}  to the 
$BRIK$ fluxes alone: this galaxy has a 
7-Gyr-old stellar population with a stellar metallicity $1.1\pm 0.2$
times solar, a stellar mass of $\log{(M_\ast/M_\odot)} = 11.00\pm
0.07$, a star 
formation rate of $\log{[(\dot{M}_\ast/(1 M_\odot \mathrm{yr^{-1}})]}
= -0.97\pm 0.35$ and a $V$ band dust extinction of $0.30 \pm 0.14$
mag.  Also plotted in the Figure is a
simple model of the emission from a dust-obscured
AGN, normalized to the 24 $\mu$m flux, to illustrate that dust can
account for the MIPS flux without contributing significant emission at
shorter wavelengths.  The curve was produced using the publicly
available 
DUSTY code for radiative transfer through dust \citep{DUSTY}, with a
simple hot-dust model after 
\citet{Gandhi_unp01}, and an input spectrum given by the mean SED for
radio-loud quasars \citep{Elvis94}.  The \emph{Chandra}
observation is also consistent with an obscured
AGN: of nine photons detected from this source seven are in the
hard (2--7 keV) X-ray band, indicating that the soft (0.5--2 keV)
X-ray emission has been absorbed.   Assuming an intrinsic X-ray
spectrum  $F_\nu \propto \nu^{-0.9}$,  the X-ray
data imply an absorption-corrected X-ray luminosity of $1.7\times
10^{42}$ erg s$^{-1}$ in the rest-frame 2--10 keV band.  Similarly,
assuming 
a radio spectrum $F_\nu \propto \nu^{-0.8}$, the 21 cm emission
translates into a rest-frame 1.4 GHz luminosity of $\rm (1.8\pm
0.6)\times 10^{23}$\,erg\,\,s$^{-1}$\,Hz$^{-1}$ (see \citealt{AEGIS20}
for details).

\section{Discussion}
\label{sec:discussion}

These observations represent
strong evidence for a dual AGN hosted by an 
early-type galaxy.  The [\ion{O}{3}] $\lambda 5007$/H$\beta$
emission-line ratio of the galaxy is iron-clad evidence for an AGN;
this is further supported by the X-ray detection.  The conclusion
that there are in fact \emph{two} accreting SMBHs in this galaxy is
driven entirely by the DEEP2 spectroscopic data.  Double peaked
emission lines like those seen in Fig.~\ref{fig:1dspec} could
be explained by three reasonable models: a rotating disk, an outflow
(or inflow), 
or a pair of orbiting emission components.  
A reasonable disk model, however, cannot easily accommodate emission
regions separated by $1.2$ kpc and $630 \kms$. Typical accretion disks
around SMBHs are smaller by several orders of magnitude and have
relativistic velocities (\emph{e.g.}, \citealt{EH03}); it is difficult
to understand how a larger disk could exhibit its brightest
AGN-like emission at distances $\sim 1$ kpc from the SMBH.
On the other hand, outflows of gas and dust are commonly invoked to
explain asymmetric [\ion{O}{3}] emission from narrow-line AGN
(\emph{e.g.},  \citealt{Heckman81}). These asymmetries, however,
typically take the form of broad, blue wings, not strong
second peaks, and in this picture we would expect one emission
component to be at rest at the host galaxy's center, unlike what is 
observed here.  
There do exist, in a few nearby Seyferts, bipolar outflows that have
spatial scales, bulk velocities, and emission-line ratios that could
explain our observations (\emph{e.g.}, \citealt{Axon98, 
  VSM01}), but these systems are quite 
kinematically disturbed, with emission components whose
velocity dispersions are comparable to
their velocity separations.  This contrasts strongly with the
relatively narrow emission-line components in this object.

We thus regard a dual SMBH as the likeliest explanation for our
observations.  The identification is quite consistent with theoretical
expectations and with previous observations. 
As discussed in the Introduction, dual SMBHs are 
expected to remain separated at $\ga 1$ kpc for $\sim 100$ Myr after a
galaxy merger and then rapidly form a parsec-scale binary 
Indeed, like this one, the previously 
observed dual AGN in NGC 6240 \citep{Komossa03}
has a separation $\sim 1$ kpc.  Unlike this one, that galaxy also
shows  
strong signatures of recent merger activity, but since such
signatures are quite subtle in dissipationless mergers,
they will not be as obvious in our 
imaging. However, the identification of a dual AGN in this galaxy
is not as secure as in NGC 6240, where \emph{Chandra} can resolve two
X-ray sources 
in the galaxies' centers; the angular scale of interest here
($0^{\prime\prime}.17$) is simply
too small to be resolved in our observations.

Nevertheless, assuming our interpretation is correct, it is possible
to infer further information about the two SMBHs. 
First, they are clearly bound within the galaxy, since their
peculiar velocities ($\sim 300 \kms$) are much smaller than the escape  
velocity ($\sim 1000 \kms$) implied by the host galaxy's measured
stellar mass.  Second, our observations  permit an estimate of the
SMBH masses.  
\citet{GH05} show that, in nearby AGN,
the width of the [\ion{O}{3}] emission lines is a reasonable proxy for 
the stellar velocity dispersion of the galaxy bulge, $\sigma_\ast$,
provided that any asymmetric wings are removed.  Our line-fitting
procedure accounts for all obvious asymmetries, so we may use
the measured linewidths to infer SMBH masses from the
$M_{\mathrm{BH}}$-$\sigma_\ast$ relation (\emph{e.g.},
\citealt{GH06}).   This relation implies $M_{\mathrm{BH}}\sim 5\times
10^5 M_\odot$ for the blueward component and $M_{\mathrm{BH}}\sim 5\times
10^6 M_\odot$ for the redward one.  By comparison, the relation
between bulge mass and SMBH mass (\emph{e.g.} \citealt{HR04}) implies
that this $M_\ast= 10^{11} M_\odot$ galaxy should harbor a BH of
mass $\sim 10^8 M_\odot$, more than an order of magnitude larger than
the total BH mass we infer.  The discrepancy is only at the $\sim 2
\sigma$ level, however, given the large scatter in the relations
discussed here.

Simple considerations might lead one to expect that
dual SMBHs reside in a relatively large fraction of galaxies: if a 
typical galaxy undergoes a major merger every few Gyr, and the
resulting dual SMBH has a lifetime of $\sim
100$ Myr, then a few percent of all galaxies---say $\sim 1000$ of the
$\sim 40000$ galaxies in DEEP2---should host such systems.  The fact
that we have 
found only one dual AGN in DEEP2 is not inconsistent with this
estimate. First, it is not clear that \emph{every} dual SMBH will exhibit AGN
activity for its full lifetime, especially from both nuclei at
once. Even if so, clean detection of double-peaked AGN emission lines
is easiest if the bright, narrow [\ion{O}{3}] emission lines 
fall in the DEEP2 wavelength range; this is true for only $\sim 30\%$ 
of the DEEP2 sample.  Also, dual AGN that occur in  
late-type galaxies may have the bulk of their optical 
emission obscured by dust (e.g., \citealt{Zakamska05}), 
so we can expect objects with observational signatures like this
one to occur mostly in early-type systems.  These constitute only
$\sim 15$\% of the DEEP2 sample.    Thus we would expect DEEP2 to have
only a few dozen objects like the one studied here if all such
objects are active.  This dual AGN was 
discovered serendipitously; a comprehensive
search of the DEEP2 catalog might find more.

We close by addressing the relatively weak emission bridging 
the two emission components in Figs.~\ref{fig:1dspec}
and~\ref{fig:2dspec}.  As already mentioned, narrow [\ion{O}{3}]
lines in  AGN spectra frequently exhibit asymmetric blue wings,
probably from outflows (\emph{e.g.}, \citealt{Heckman81}).  If such an
outflow is present for the redward AGN here, it could explain the
bridging emission. Given the unusual nature of this galaxy's spectrum,
however, this explanation is by no means certain.
Infrared imaging and integral-field unit observations on an 8-meter
class telescope with adaptive optics, or narrow-band imaging of the
[\ion{O}{3}] line from space, could help to
further characterize this emission and the system as a whole.

\begin{acknowledgments}
We thank M. Boylan-Kolchin, E. Quataert and the anonymous referee 
for helpful comments.  
This work is based in part on observations made with the Spitzer
Space Telescope, which is operated by the Jet Propulsion Laboratory,
California Institute of Technology under a contract with NASA. Support
for this work was provided by NASA through an award issued by
JPL/Caltech.
The National Radio Astronomy Observatory is operated by Associated
Universities, Inc. under a cooperative agreement with the National
Science Foundation.
Further relevant acknowledgments for this work are in \citet{AEGIS} 

\end{acknowledgments}


\begin{thebibliography}{}

\bibitem[Axon {et~al.}(1998)Axon {\em et~al.}]{Axon98}
Axon, D.~J., et~al. 1998, ApJ, 496, L75

\bibitem[Ballo {et~al.}(2004)Ballo {\em et~al.}]{Ballo04}
Ballo, L., et~al. 2004, ApJ, 600, 634

\bibitem[Begelman, Blandford, \& Rees(1980)Begelman, Blandford, and
  Rees]{BBR80}
Begelman, M.~C., Blandford, R.~D., \& Rees, M.~J. 1980, Nature, 287, 307

\bibitem[Bruzual \& Charlot(2003)Bruzual and Charlot]{BC03}
Bruzual, G., \& Charlot, S. 2003, MNRAS, 344, 1000

\bibitem[Davis {et~al.}(2006)Davis {\em et~al.}]{AEGIS}
Davis, M., et~al. 2006, ApJ, in prep

\bibitem[Elvis {et~al.}(1994)Elvis {\em et~al.}]{Elvis94}
Elvis, M., et~al. 1994, ApJS, 95, 1

\bibitem[Eracleous \& Halpern(2003)Eracleous and Halpern]{EH03}
Eracleous, M., \& Halpern, J.~P. 2003, ApJ, 599, 886

\bibitem[Faber {et~al.}(2005)Faber {\em et~al.}]{Faber05}
Faber, S.~M., et~al. 2005, ApJ, submitted (astro-ph/0506044)

\bibitem[Gandhi {et~al.}(2001)Gandhi {\em et~al.}]{Gandhi_unp01}
Gandhi, P., et~al. 2001, (astro-ph/0106139)

\bibitem[Greene \& Ho(2005)Greene and Ho]{GH05}
Greene, J.~E., \& Ho, L.~C. 2005, ApJ, 627, 721

\bibitem[Greene \& Ho(2006)Greene and Ho]{GH06}
Greene, J.~E., \& Ho, L.~C. 2006, ApJ, 641, L21

\bibitem[Guainazzi {et~al.}(2005)Guainazzi {\em et~al.}]{Guainazzi05}
Guainazzi, M. et~al. 2005, A\&A, 429, 9

\bibitem[H\"{a}ring \& Rix(2004)H\"{a}ring and Rix]{HR04}
H\"{a}ring, N., \& Rix, H.-W. 2004, ApJ, 604, L89

\bibitem[Heckman {et~al.}(1981)Heckman {\em et~al.}]{Heckman81}
Heckman, T.~M., et~al. 1981, ApJ, 247, 403

\bibitem[Hudson {et~al.}(2006)Hudson {\em et~al.}]{Hudson06}
Hudson, D.~S., et~al. 2006, A\&A, in press (astro-ph/0603272)

\bibitem[Ivezic, Nenkova, \& Elitzur(1999)Ivezic, Nenkova, and Elitzur]{DUSTY}
Ivezic, Z., Nenkova, M., \& Elitzur, M. 1999, (astro-ph/9910475)

\bibitem[Ivison {et~al.}(2006)Ivison {\em et~al.}]{AEGIS20}
Ivison, R.~J., et~al. 2006, ApJ, in prep.

\bibitem[Junkkarinen {et~al.}(2001)Junkkarinen {\em et~al.}]{Junkkarinen01}
Junkkarinen, V., et~al. 2001, ApJ, 549, L155

\bibitem[Kauffman(2003)Kauffman]{Kauffmann03}
Kauffman, G. 2003, MNRAS, p. 1055

\bibitem[Kochanek, Falco \& Mu\~{n}oz(1999)Kochanek, Falco and Mu\~{n}oz]{Kochanek99}
Kochanek, C.~S., Falco, E.~E. and Mu\~{n}oz, J.~A. 1999, ApJ, 510, 590

\bibitem[Komossa {et~al.}(2003)Komossa {\em et~al.}]{Komossa03}
Komossa, S., et~al. 2003, ApJ, 582, L15

\bibitem[Lotz {et~al.}(2006)Lotz {\em et~al.}]{Lotz06}
Lotz, J., et~al. 2006, ApJ, submitted (astro-ph/0602088)

\bibitem[Maness {et~al.}(2004)Maness {\em et~al.}]{Maness04}
Maness, H.~L., et~al. 2004, ApJ, 602, 123

\bibitem[Merritt \& Milosavljevi\'{c}(2005)Merritt and Milosavljevi\'{c}]{MM05}
Merritt, D., \& Milosavljevi\'{c}, M. 2005, Living Rev. Relativity, 8, 8,
  (astro-ph/0410364)

\bibitem[Milosavljevi\'{c} {et~al.}(2002)Milosavljevi\'{c} {\em
  et~al.}]{MMRv02}
Milosavljevi\'{c}, M., et~al. 2002, MNRAS, 331, L51

\bibitem[Nandra {et~al.}(2006)Nandra {\em et~al.}]{Nandra06}
Nandra, K., et~al. 2006, ApJ, in prep.

\bibitem[Peng {et~al.}(2002)Peng {\em et~al.}]{Galfit}
Peng, C.~Y., et~al. 2002, AJ, 124, 266

\bibitem[Rodriguez {et~al.}(2006)Rodriguez {\em et~al.}]{Rodriguez06}
Rodriguez, C., et~al. 2006, ApJ, in press (astro-ph/0604042)

\bibitem[Salim {et~al.}(2005)Salim {\em et~al.}]{Salim05}
Salim, S., et~al. 2005, ApJ, 619, L39

\bibitem[S\'{e}rsic(1968)S\'{e}rsic]{Sersic68}
S\'{e}rsic, J.~L. 1968.
\newblock Atlas de Galaxias Australes, C\'{o}rdoba: Obs. Astron., Univ. Nac.
  C\'{o}rdoba

\bibitem[van Dokkum {et~al.}(2005)van Dokkum {\em et~al.}]{vanD05}
van Dokkum, P., et~al. 2005, AJ, 130, 2647

\bibitem[Veilleux, Shopbell, \& Miller(2001)Veilleux, Shopbell, and
  Miller]{VSM01}
Veilleux, S., Shopbell, P.~L., \& Miller, S.~T. 2001, AJ, 121, 198

\bibitem[Willmer {et~al.}(2005)Willmer {\em et~al.}]{Willmer05}
Willmer, C. N.~A., et~al. 2005, ApJ, in press

\bibitem[Yan {et~al.}(2006)Yan {\em et~al.}]{Yan06}
Yan, R., et~al. 2006, ApJ, in press. (astro-ph/0512446)

\bibitem[Zakamska {et~al.}(2005)Zakamska {\em et~al.}]{Zakamska05}
Zakamska, N.~L., et~al. 2005, AJ, 129, 1212

\end{thebibliography}
\end{document}